\documentclass[iopart,twocolumn,amsmath,amssymb]{revtex4-1}

\usepackage{graphicx}
\usepackage{bm}
\usepackage{gensymb}
\usepackage{floatrow}
\usepackage{amsmath}
\usepackage{color}
\usepackage[T1]{fontenc}

\DeclareGraphicsExtensions{.eps,.jpg,.pdf}

\begin{document}

\title{Mesoscopic quantum effects in a bad metal, hydrogen-doped vanadium dioxide}

\author{Will J. Hardy}
\affiliation{Applied Physics Graduate Program, Smalley-Curl Institute, Rice University, 6100 Main St., Houston, Texas 77005, USA}
\author{Heng Ji}
\affiliation{Department of Physics and Astronomy, Rice University, 6100 Main St., Houston, Texas 77005, USA}
\author{Hanjong Paik}
\affiliation{Department of Materials Science and Engineering, Cornell University, Ithaca, New York 14853, USA}
\author{Darrell G. Schlom}
\affiliation{Department of Materials Science and Engineering, Cornell University, Ithaca, New York 14853, USA}
\affiliation{Kavli Institute at Cornell for Nanoscale Science, Ithaca, New York 14853, USA}
\author{Douglas Natelson$^*$}
\affiliation{Department of Physics and Astronomy, Rice University, 6100 Main St., Houston, Texas 77005, USA}
\affiliation{Department of Electrical and Computer Engineering, Rice University, 6100 Main St., Houston, Texas 77005, USA}
\affiliation{Department of Materials Science and Nanoengineering, Rice University, 6100 Main St., Houston, Texas 77005, USA 
\\$^*$e-mail: natelson@rice.edu}

\date{\today}

\begin{abstract}
The standard treatment of quantum corrections to semiclassical electronic conduction assumes that charge carriers propagate many wavelengths between scattering events, and succeeds in explaining multiple phenomena (weak localization magnetoresistance (WLMR), universal conductance fluctuations, Aharonov-Bohm oscillations) observed in polycrystalline metals and doped semiconductors in various dimensionalities.  We report apparent WLMR and conductance fluctuations in H$_{x}$VO$_{2}$, a poor metal (in violation of the Mott-Ioffe-Regel limit) stabilized by the suppression of the VO$_{2}$ metal-insulator transition through atomic hydrogen doping.  Epitaxial thin films, single-crystal nanobeams, and nanosheets show similar phenomenology, though the details of the apparent WLMR seem to depend on the combined effects of the strain environment and presumed doping level.  Self-consistent quantitative analysis of the WLMR is challenging given this and the high resistivity of the material
 , since the quantitative expressions for WLMR are derived assuming good metallicity.  These observations raise the issue of how to assess and analyze mesoscopic quantum effects in poor metals.

\end{abstract}

\maketitle

\section{Introduction}

Since the late 1970s, many insights have been gained into the effects of quantum coherence and quantum interference on electronic conduction in ordinary metals and semiconductors\cite{Imry2008}.  Conduction is influenced by the interference between trajectories as charge carriers scatter from static disorder, while inelastic scattering processes suppress coherence on the scale of the coherence length, $L_{\phi}$.  Magnetic flux alters relative phases of various trajectories \textit{via} the Aharonov-Bohm phase.  Magnetoresistive effects (\textit{e.g.}, weak localization\cite{Bergmann1984}, universal conductance fluctuations\cite{Lee1985}) are then experimental means of studying the evolution of quantum coherence as a function of control parameters such as temperature, dimensionality, disorder, and carrier density.  The standard theoretical treatment of these effects is perturbative in $(1/k_{\mathrm{F}}\ell)$, where $k_{\mathrm{F}}$ is the Fermi wavelength and $\ell$ is the c
 arrier mean free path, so that carriers propagate many wavelengths between scattering events, a hallmark of a ``good metal.''

In contrast, some materials are ``bad metals,'' such that $k_{\mathrm{F}}\ell$ as inferred from the resistivity is $<$ 1, or roughly equivalently, when the Mott-Ioffe-Regel limit is violated\cite{Gunnarsson2003,Hussey2004}.  This situation most commonly arises at high temperatures in correlated materials.  Conventionally, when scattering is dominated by disorder, materials with small $k_{\mathrm{F}}\ell$ cross over into the strongly localized regime as $T$ is lowered\cite{Khavin1998}.  Quantum corrections to conduction in bad metals are comparatively unexplored.

VO$_{2}$ is a strongly correlated material with an insulator-to-metal phase transition at about 340 K.  Its conductivity can increase by up to 5 orders of magnitude when the material is heated across the transition temperature and the material transforms from a low temperature, monoclinic, insulating phase to a high temperature, rutile, conducting phase\cite{Ladd1969}. This metallic state is a moderately correlated bad metal\cite{Qazilbash2006} that has not been studied previously at low temperatures due to the transition to the insulator. Atomic hydrogen can be doped into VO$_{2}$, either with\cite{Wei2012,Filinchuk2014} or without catalyst\cite{Lin2014}, and this doping dramatically affects the properties of the material.  Based on the neutron powder diffraction measurements, over a broad range of concentrations of the intercalated hydrogen atoms ($x$ in H$_{x}$VO$_{2}$), new orthorhombic phases are stabilized, with crystal structures similar to the pristine tetragonal ruti
 le structure except that $a$-axis lattice parameter is slightly different from that of the $b$-axis.  These new phases possess electronic properties similar to the pristine rutile phase in terms of resistivity (and theory anticipates their metallic behavior), and can be cooled down to 2~K without any phase transition.  Here, we find that in hydrogen-doped nanobeams and micron-scale flakes, the resistivity is only weakly dependent on the hydrogen concentration (as inferred from varying the hydrogen exposure time), consistent with the metallicity appearing as the result of thermodynamic phase stabilization\cite{Filinchuk2014} rather than traditional doping. A different situation is observed in thin films, however, where instead of reaching a stabilized conducting state that does not change appreciably with further hydrogenation, the films initially become more conducting upon hydrogen exposure, then trend more insulating with longer exposure times. This could presumably occur 
 if the doping fraction $x \rightarrow$ 1 (stoichiometric HVO$_2$, which is reported to have completely filled 3$d$ orbitals, and thus no available states at the Fermi level), as was  recently observed in a study on reversible hydrogen doping in thin VO$_2$ films\cite{Yoon2016}), or it may be a consequence of the different strain situation for films compared to bulk VO$_2$ or micro-crystals. No systematic experimental study has been done of the low-temperature electronic transport properties of hydrogen-doped VO$_2$, in the case of nanobeams, micro-crystals, or thin films.

  In this work, we report the low temperature transport properties of  hydrogenated VO$_{2}$. Three types of single-crystal samples have been used: epitaxial thin films on single-crystal TiO$_{2}$ substrates (the rutile polymorph), nanobeams, and micron-scale flake-like crystals. Hall measurements on the pristine metallic film samples at room temperature, as well as on the doped films at various temperatures, show a very small Hall signal, similar to that reported previously at room temperature and above on pristine, metallic VO$_{2}$\cite{Rosevear1973,Ruzmetov2009}.  The very small Hall response in the doped material implies that multiple carrier types are likely present in the metallic state. After hydrogenation, we observe abundant magnetoresistance (MR) responses in these samples at low temperatures.  For thin films of 10 nm thickness, an overall high-field positive MR is observed at low temperatures, while the low-field MR may have either a peak or dip at helium temperat
 ures.   The nanobeam and micron flakes also exhibit positive MR of increasing magnitude as $T$ is reduced below 20~K, independent of field direction, qualitatively similar to a 3d weak anti-localization response. Mesoscale samples show conductance fluctuations as a function of magnetic field.  Interpretation of these results is discussed. 

\section{Experimental Techniques}

Three kinds of single-crystal VO$_{2}$ samples were used for hydrogen doping and transport measurements.  Epitaxial VO$_{2}$ thin films were grown on rutile (001) TiO$_{2}$ single-crystal substrates by reactive molecular-beam epitaxy (MBE) in a Veeco GEN10.  A 23\% HF aqueous solution was used to reduce the metal impurity contamination on the (001) TiO$_{2}$ surface before growth \cite{Paik2015}. Initially, films of 30~nm thickness were grown, but were found to be extremely susceptible to microcracking during temperature cycling and cutting the as-grown samples into smaller pieces, as observed recently by Paik $et~al$. \cite{Paik2015} To avoid this issue, 10 nm thick films were grown in the same manner (three nominally identical samples denoted HP1703, HP1704, and HP1706), and measured after minimal processing (contact deposition only). Through a shadow mask, we used an electron-beam evaporator to deposit six contacts (5~nm vanadium followed by 35 or 50~nm gold) along the sid
 e of 1~cm $\times$ 1~cm VO$_{2}$ films to form Hall bar structures. The other two sample types were VO$_{2}$ nanobeams and micron-scale flake-like single crystals, both grown by physical vapor deposition (PVD) on Si/SiO$_{2}$ substrates using methods reported previously,\cite{Wei2009,Wei2012} with wire or flake morphology determined by the position of the substrate within the growth furnace. The nanobeams could be up to 100~$\mu$m in length, and the widths and thicknesses ranged from 50~nm to 1~$\mu$m.  For the flakes, the thickness was between 50 \textendash 100~nm, and the dimensions were tens of $\mu$m by tens of $\mu$m.  Usually, both wires and flakes could be achieved on one large substrate during the same growth, improving the likelihood that they would have similar properties.  We used electron-beam lithography to define the contacts for 4-terminal transport measurements, and deposited 5~nm vanadium followed by 35~nm gold as contacts.  We intentionally chose wires and
  flakes less than 200~nm thick to avoid difficulties in contact continuity at the edges of thick samples.

After an initial temperature-dependent resistivity measurement (in the range 200--400~K), all test structures were then treated by a catalyst-free atomic hydrogenation process.\cite{Lin2014}  In a dedicated tube furnace, hydrogen gas was flowed into the tube at rate 100~cc/min; a tungsten filament heated by a 45~W power supply was located in the middle of the tube, and samples to be hydrogenated sat 1~cm away from the filament on the downstream side.  A valve on the pumping side was adjusted to stabilize the tube pressure at 10~Torr.  During this process, the hot filament splits molecular hydrogen gas at some rate, and as observed previously the hydrogen atoms then diffuse into rutile VO$_{2}$ along its $c$-axis, without the need for any catalyst (diffusion in the rutile structure is much faster than in the monoclinic phase).  Because the VO$_{2}$ film on the TiO$_{2}$ substrate has its metal-insulator phase transition shifted to below room temperature, due to its high strain
  from lattice mismatch with the substrate, the atomic hydrogenation process could happen at room temperature. Since the $c$-axis of the film was perpendicular to the substrate, the doping process took no more than 30 seconds to complete for the 10~nm thickness (2 min for the 30 nm thickness), with hydrogen entering the film evenly from the whole surface, resulting in apparently homogeneous doping (based on the uniform appearance and color of the treated films, although we later discuss evidence for local nonuniformity and time variation of uniformity of the doping). Accurate determination of the hydrogen concentration is challenging and requires specialized analytical techniques sensitive to light elements, such as annular brightfield scanning transmission electron microscopy (ABF-STEM), elastic recoil detection (ERD),\cite{Yoon2016} or highly destructive techniques involving chemical decomposition of the sample and analysis of the reaction products,\cite{Horiba} which are b
 eyond the scope of this work.  For the films, it is possible that there would be some modification of the strain at the interface between the VO$_2$ and TiO$_2$ after doping, as suggested by previous work that shows changes in the unit cell parameters with hydrogen incorporation\cite{Filinchuk2014}. This could perhaps be assessed with appropriate x-ray scattering techniques before and after hydrogenation.  Given the large pre-existing strain due to the TiO$_2$/VO$_2$ lattice mismatch, however, we believe that any such modification of strain is likely small compared to the pre-existing strain.

The first film sample, denoted HP1703, was exposed to atomic hydrogen for an initial time of 2 minutes, followed by a second treatment for an additional 2 minutes, and showed evidence of doping beyond the point of stabilizing the metallic phase. After the initial 2-minute exposure, the sample shows a quite insulating temperature dependence of resistivity upon cooling to 75~K, becoming even more so after the second 2-minute atomic hydrogen exposure ($>$~2000 m$\Omega$-cm; see Figure S1). This is consistent with a recent report that epitaxial VO$_{2}$ films doped with hydrogen by the catalytic spillover method first become more conducting at low doping levels, but with progressively more doping the trend reverses, resulting in substantially insulating behavior as the stoichiometry approaches HVO$_{2}$.\cite{Yoon2016} Further, it would imply that the doping fraction in this film is large ($x$ $\rightarrow$ 1) based on comparison with prior diffraction studies of hydrogenated VO$
 _2$ powder, where an increasing hydrogen doping fraction (up to $x$ $\sim$ 0.6) is accompanied by stabilization of two successive orthorhombic crystal phases, which are most similar to the pristine metallic state.\cite{Filinchuk2014} Since the goal of this work was to stabilize a relatively conducting (but still badly metallic) phase down to low temperatures, two subsequently measured samples, HP1704 and HP1706, were doped for only 30~s. These remain relatively conducting over the temperature range 2--300~K (though with resistivity magnitudes well above the regime of good metal behavior). Despite the fact that they have the same geometry (effective number of squares) and were both doped for the same nominal time, their post-hydrogenation temperature-dependent resistivity curves are substantially different from each other,    by a factor of $\sim$ 2.7$\times$ at room temperature and by $\sim$ 6$\times$ at 2~K (as shown in Figs. 1a and 2a). This may imply an extreme sensitivit
 y to the exact doping level, with a crucial dependence on the treatment process details (a difference of a few seconds in doping time, or  the exact sample-filament distance). The built-in strain of the films may also be relevant, in light of this apparently enhanced sensitivity to hydrogen exposure time compared with powder or micro-crystal samples.

 For hydrogenation of the wires and flakes, the furnace had to be turned on to increase the ambient temperature to 425~K, so that the VO$_{2}$ crystals had the rutile structure best suited for hydrogenation.  Since the $c_{\mathrm{R}}$-axis for these crystals was along the long dimension of the wires or along an in-plane dimension of the flakes, it took 15 minutes for the VO$_{2}$ samples to be fully hydrogenated (in the sense that samples appeared homogeneous in color after cooling to room temperature, and showed no significant changes in measured electronic properties if hydrogenated for longer periods).  While hydrogen content within the samples is not readily quantifiable, again we point out that for PVD-grown crystals, the measured electronic properties appear to be fairly independent of hydrogen content above the threshold required to stabilize the orthorhombic phase down to low temperatures, perhaps due to the relatively slow hydrogen diffusion compared to the film geo
 metry. After atomic hydrogen treatment, the overall trends of the transport properties are highly reproducible from one sample to another: Suppression of the metal-insulator transition, stabilization of a conducting but badly metallic phase down to liquid helium temperatures, with a very flat temperature dependence to the resistivity; evidence of conducting phase stabilization rather than semiconductor-like simple carrier doping; and absence of low-temperature strong localization. There is some relatively minor sample-to-sample variability, visible in the detailed shape of the resistivity and MR curves, which we believe results from a combination of the exact hydrogen doping fraction, the possibly non-uniform distribution of hydrogen within the VO$_2$ lattice, and the unique strain environment for each sample.

Although the main focus of this work is on the intriguing transport properties of a bad metal where $k_{\mathrm{F}}\ell < $1 that nevertheless remains relatively conductive down to low temperatures, it is worth pointing out here that hydrogenated VO$_2$ may also lend itself to practical applications, though not without overcoming certain technical obstacles. One natural possibility would be to use VO$_2$ as a hydrogen sensor. Challenges include the fact that hydrogen uptake is most favored at high temperatures, along particular crystallographic directions, and when VO$_2$ is already in its metallic phase; at low temperatures, hydrogen uptake in pristine material is many times slower than in the metallic phase, so detection speed may be slow.\cite{Lin2014} Electrical switching is another possible option, $e.g.$, using current or optical heating to locally de-hydrogenate a region of metallic hydrogenated VO$_2$ ($e.g.$, one nanowire in an array of wires all initially prepared i
 n the hydrogenated state), driving it to the insulating phase after cooling to room temperature. An array of such switches could work as a memory. The reset process (by re-hydrogenation) could present practical limitations.

All the transport measurements were done in a Quantum Design Physical Property Measurement System (PPMS), with a base temperature of 1.8~K, and magnetic field up to 9~T.  A typical 4-terminal measurement setup was applied for all temperature-dependent resistivity and magnetoresistance (MR) measurements.  A low frequency (7\textendash 17~Hz) AC signal was powered and collected by two lock-in amplifiers (Signal Recovery model 7270 and model 7265), one current pre-amplifier (Stanford Research Systems model SR570), and one voltage pre-amplifier (Stanford Research Systems model SR560).  For Hall measurements, one additional voltage pre-amplifier was used.  For all low temperature measurements, currents were limited to below 1~$\mu$A to minimize Joule heating.

\begin{figure*}[!htb]
\includegraphics[width=\textwidth]{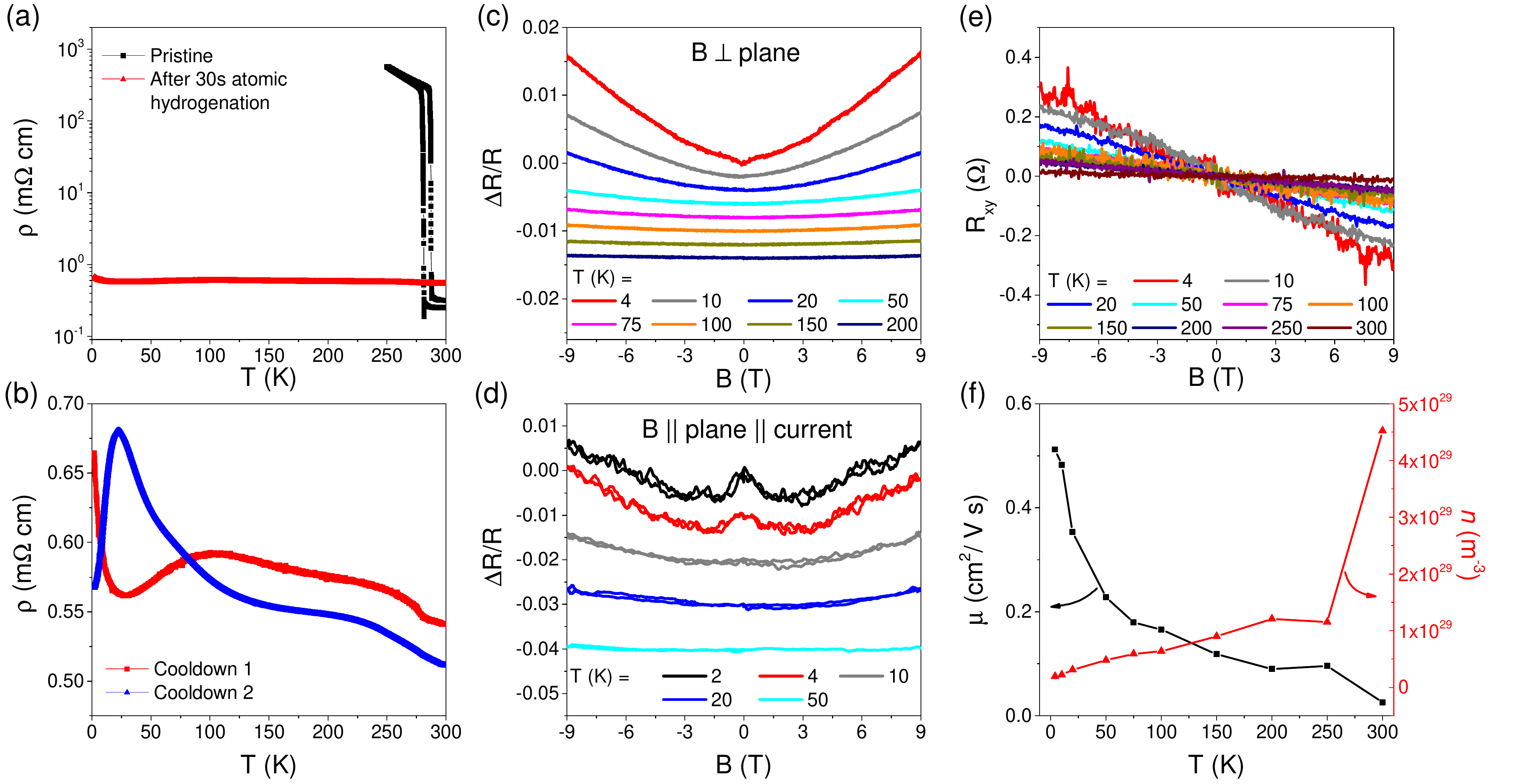}
\caption{(a) Temperature dependence of the resistivity of the VO$_{2}$ film sample HP1704 (10~nm thickness), before (black squares) and after (red triangles) the 30~s atomic hydrogenation process. The metal-insulator transition in the pristine material is marked by a sharp increase of resistivity of \textgreater ~3 orders of magnitude at $\sim$ 282~K on cooling, whereas after hydrogenation, the phase transition disappears and the resisitivity is nearly flat with temperature. (b) A closer view of the resistivity immediately after hydrogenation (red squares) and after a subsequent warming and cooling cycle (blue triangles), showing that the detailed shape of the curve changes over time, even though the sample remains in the controlled cryostat environment, perhaps due to time evolution of the hydrogen distribution in the lattice. (c,d) Hydrogenated film MR curves at various temperatures with the field (c) perpendicular or (d) parallel to the substrate and current direction. Cur
 ves are offset by (c) -0.002 or (d) -0.01  for clarity. An overall high-field  positive MR develops as the temperature is reduced below $\sim$ 150~K, growing as the temperature decreases. A small dip (perpendicular case) or peak (parallel case) appears at 4~K and below, likely due to localization effects. (e) Hall resistance of the hydrogenated film measured at various temperatures. The data have been antisymmetrized to remove a longitudinal component artifact. (f) Calculated mobility (left axis) and carrier density (right axis) of the hydrogenated film inferred from resisitvity and Hall data assuming a single carrier type ($n$-type). The unphysically large inferred carrier density implies that more than one carrier type likely contributes to transport in the hydrogenated film. 
 \label{fig:figure1}}
\end{figure*}

\section{Results and Discussion} 

\subsection{VO$_{2}$ film}
Before hydrogenation, the VO$_{2}$ films of 10~nm thickness show a typical sharp phase transition behavior (Fig. 1a) with the transition temperature at $\sim$ 282~K on cooling (290~K for the 30 nm film), instead of the 340~K transition of unstrained bulk VO$_{2}$.  This reduced transition temperature has been reported previously, and was well explained by a compressive strain along the $c_{\mathrm{R}}$-axis\cite{Muraoka2002}.  After the 30s room-temperature atomic hydrogenation process (samples HP1704 and HP1706), the films remain in a conducting state with a resistivity similar to that of the pristine metallic state.  Unlike a good metal, the temperature dependence of resistivity of the films shows an overall negative slope, with behavior below 100~K that changes after subsequently warming up and cooling down again (Fig. 1b). The different response on the subsequent cooldown may be due to redistribution of hydrogen concentration in the lattice over time.  The total change of
  resistivity from 300~K to 2~K is no more than a few times for both 10~nm films hydrogenated for 30~s (or less than an order of magnitude for the 30~nm film). There is no obvious sign of impending strong localization at low temperatures, in contrast to the case of very strongly  disordered materials ($e.g.$, the doped semiconductor wires of Gershenson $et~al.$\cite{Gershenson1997}), where a crossover to strong localization is observed as $k_{\mathrm{F}}\ell$ approaches 1.

Even at room temperature, the resistivity of the hydrogenated VO$_{2}$ film is much higher than expected for a typical metal.  Based on the Mott-Ioffe-Regel limit (MIRL), the semiclassical transport model for quasiparticles is only applicable when the mean free path $\ell$ is larger than the lattice constant $a$, leading to a criterion of $k_{\mathrm{F}}\ell \sim 1$ (or $2\pi$). This leads to a maximum resistivity of metals in the conventional model on the order of $a\hbar/e^{2}$.  When considering VO$_{2}$ with a lattice constant of 2.85~\AA~ along the $c_{\mathrm{R}}$-axis\cite{Qazilbash2006},  its maximum resistivity should then be around 117~$\mu\Omega$-cm.  In our measurements, however, the resistivity of the hydrogenated VO$_{2}$ film of 10~nm thickness is larger than 540~$\mu\Omega$-cm for one sample (HP1704) and larger than 1450~$\mu\Omega$-cm for a second sample (HP1706) (or 700~$\mu\Omega$-cm for the 30~nm film) throughout the temperature range from 300~K to 2~K,  w
 ith similar values measured for PVD-grown wires and flakes, implying the inadequacy of semiclassical quasiparticle conduction in VO$_{2}$.  The violation of MIRL has been observed at high temperature in some transition metal oxides, like VO$_{2}$\cite{Qazilbash2006}, and some correlated superconducting oxides, like Sr$_{2}$RuO$_{4}$ \cite{Tyler1998}, La$_{x}$Sr$_{1-x}$CuO$_{4}$, and YBa$_{2}$Cu$_{3}$O$_{7}$\cite{Gurvitch1987}.   There are few reports about violation of the MIRL near the threshold of strong localization at low temperature\cite{Valla2002,Son2010,Scherwitzl2011}.

We performed MR measurements on the VO$_{2}$ films (samples HP1704, HP1706) at various temperatures. For sample HP1704, which appears to be less doped due to its smaller increase of resistivity with temperature, the MR is shown in Fig. 1 with the external magnetic field either perpendicular (Fig. 1c) or parallel (Fig. 1d) to the film surface and current  direction. The relative change of resistance $\frac{\Delta R}{R_{0}}$ =  $\frac{R(B)-R_{0}}{R_{0}}$ is plotted as a function of external field strength. In both orientations, an overall high-field MR develops and grows as the temperature is reduced below $\sim$ 150~K, with a maximum magnitude less than 2\% at low temperatures, and a slope that shows no sign of saturation by 9~T. At the lowest temperatures (2 or 4~K), the low-field region of the MR curves shows a dip when the field is perpendicular to the plane, or a peak when the field is parallel to the plane. Film HP1706, which despite nominally identical processing appears
  to be more strongly doped (based on its larger room-temperature magnitude of resistivity after doping and steeper increase of resistivity with falling temperature, shown in Figure 2(a)), also shows a high-field positive MR of similar magnitude at temperatures below $\sim$ 150~K, and has a zero-field peak at temperatures below 4~K  with the field perpendicular to plane (Figure 2(b)).

\begin{figure}
\includegraphics[width=\textwidth]{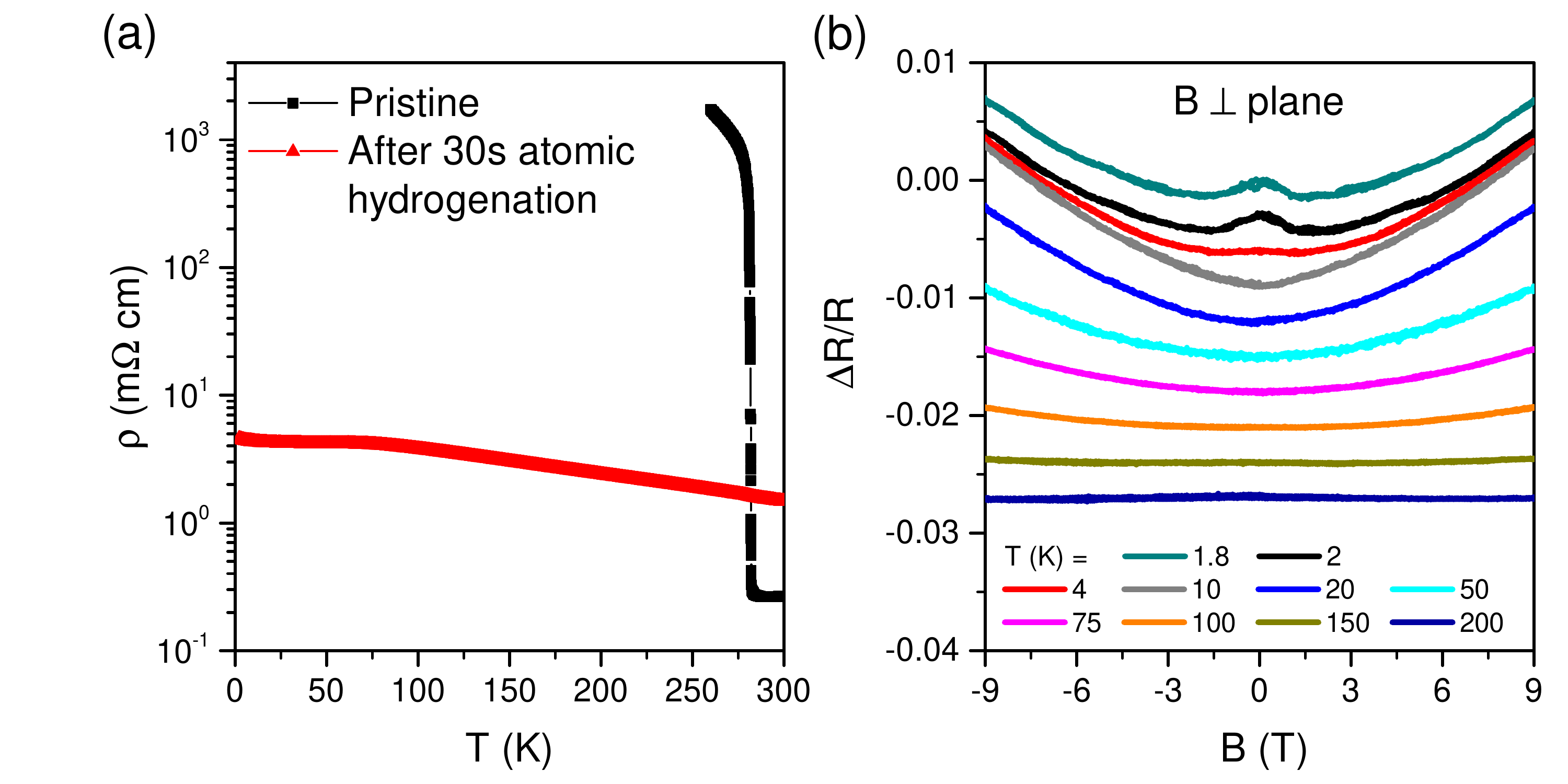}
\caption{ Transport measurements of the 10 nm thick film HP1706. (a) Resistivity before (black curve, cooling data) and after (red curve) 30~s atomic hydrogenation. (b) MR curves of the hydrogenated film at various temperatures, with the field perpendicular to the plane of the sample. The curves are offset by -0.003 per temperature value for clarity. An overall positive sign of MR develops below $\sim$ 150~K, and a zero-field peak emerges when cooling below 4~K.  \label{fig:figure2}}
\end{figure}

 This magnitude of the magnetoresistance, its functional form, and its suppression with increasing temperature, resemble the predictions of the theory of weak localization\cite{Bergmann1984}.  Weak localization usually occurs at low temperatures when the inelastic scattering time ($\tau_{\mathrm{i}}$) is much larger than the elastic scattering time ($\tau_{\mathrm{e}}$), so that the electrons can experience momentum relaxation and diffusion before loss of phase coherence.  In the absence of strong spin-orbit coupling, trajectories corresponding to closed loops and their time-reversed partners interfere constructively, favoring back-scattering, i.e., localization, and result in a slightly larger resistance.  When a magnetic field is applied, the constructive interference is broken due to the Aharonov-Bohm phase, suppressing the back-scattering and leading to negative MR.  In a system with strong spin-orbit scattering, this interference at the origin point becomes destructive, 
 suppressing back-scattering; the MR sign flips to positive, producing weak anti-localization.  The functional form of the MR depends on the effective dimensionality of the sample with respect to the coherence length. In conventional (good) metal systems, WL or WAL data can be modeled by functional forms that are tailored for the case of either two \cite{Hikami1980} or three \cite{Baxter1989} dimensions, and the relevant field scales that allow determination of the coherence length, spin-orbit contribution, spin-flip contribution, and  inelastic scattering can be deduced from the fitting parameters. Such is the case in ``dirty'' thin films of noble metals -- while disordered and thus subject to weak localization, they remain in the ``good'' metal limit, with $k_{\mathrm{F}}\ell >>$ 1. \cite{Bergmann1984}
 
 On the other hand, quantitatively fitting the hydrogenated VO$_2$ data to the theoretical expressions for WL/WAL with self-consistent values of parameters, particularly the role of elastic scattering, is not possible.  The MIR-violating resistivity and difficulty extracting meaningful carrier densities from Hall data are the source of the problems.  We can use the Einstein relation, $\sigma = e^{2}D \nu(E_{\mathrm{F}})$, where $\sigma = 1/\rho$, $e$ is the electronic charge, and $\nu(E_{\mathrm{F}})$ is the density of states at the Fermi level, to infer a value for $D$.  Using the estimate of $k_{\mathrm{F}} \sim 10^{8}$ cm$^{-1}$ employed by Qazilbash \textit{et al.}\cite{Qazilbash2006} and assuming a spherical Fermi surface and an effective mass close to the free electron mass, one finds $\nu(E_{\mathrm{F}}) \sim  8 \times 10^{46}$ J$^{-1}$m$^{-3}$, implying a diffusion constant $D \sim  5 \times 10^{-5}$ m$^{2}$/s when $\rho = 1$ m$\Omega$-cm.  As expected for the MIRL si
 tuation, however, this also implies a mean free path small compared to the lattice spacing and is therefore likely unphysical.  We note that employing a different value of the effective mass $m^*$ could also affect the inferred $D$, with increasing $m^*$ causing a reduction of $D$. Early works \cite{Berglund1969} estimate the effective mass at $1.6 - 7~m_0$, depending on the method, while more recent analyses based on measurements of optical conductivity put the effective mass at no more than a few times the free electron mass.  For example, the work by Okazaki \textit{et al.} \cite{Okazaki2006} suggests $m^*$ $\sim$ 4 $m_0$, while recent theory work\cite{Belozerov2011} suggests $m^* \sim 2~m_{0}$. Further, optical conductivity results indicate the possibility of a diverging effective mass as the system nears the metal-insulator transition \cite{Qazilbash2006}, which would imply a vanishing diffusion constant. As discussed below in the context of the Hall number, the electro
 nic structure of metallic VO$_{2}$ remains a topic of debate\cite{Saeki2009}.

These issues with $D$ and $k_{\mathrm{F}}\ell$ make it very difficult to apply the quantitative formula of WAL/WL meaningfully, particularly as the underlying theory is based on the assumption that $k_{\mathrm{F}}\ell >> 1$.  However, the qualitative features of the MR (growing magnitude and increasing ``cuspiness'' near zero field as $T\rightarrow 0$) suggest that the MR mechanism is analogous to weak localization. To support this, we also present the film sample MR data plotted instead as magnetoconductance, in units of $e^2/h$ (see Fig. S2). The maximum magnitude change of conductance over the $\pm$ 9~T field range is $\sim$ 1.9~$e^2/h$ (sample HP1704, field in plane), $\sim$ 2.5~$e^2/h$ (HP1704, field out of plane), or $\sim$ 0.28~$e^2/h$ (sample HP1706, field out of plane), values which are of the typical magnitude of $WL/WAL$.  Also, the similar high-field MR for in-plane and out-of-plane field orientations imply that the system is effectively in or close to the 3d limi
 t, with a coherence length comparable to film thickness.  These observations show the need for treatments of quantum corrections to conduction applicable to such bad metals.

We note that an isotropic negative magnetoresistance quantitatively similar in magnitude and field scale to these observations has been reported previously\cite{Scherwitzl2011} in few-unit-cell-thick films of the correlated material LaNiO$_{3}$.  In that system this effect was interpreted as evidence for the importance of magnetic fluctuations\cite{Maekawa1981} acting as a source of inelastic scattering, and possible proximity to a spin glass state\cite{Nigam1983} or antiferromagnetic ordering. While the partially filled vanadium $3d$ band could in principle be relevant for magnetism in H$_{x}$VO$_{2}$, measurements to date have found no evidence of magnetic ordering\cite{Filinchuk2014}.  

Given the substantial change in the shape of the resistivity versus temperature curve for successive cooldowns of the HP1704 film sample, and the larger resistivity of sample HP1706 after doping with nominally the same hydrogenation protocol as for HP1704, it is worth considering whether the presence of either a dip or peak in the low-field MR is a consequence of both the hydrogen doping fraction and the redistribution of hydrogen within the lattice over time, rather than an intrinsic anisotropic response of the material. The absence of anisotropy is further supported by measurements on nanobeam and micron-scale flake samples, as we describe in a subsequent section.

We also performed Hall measurements on hydrogenated 10~nm thickness VO$_{2}$ films.  In the metallic state of pristine VO$_{2}$ single crystals\cite{Rosevear1973} and thin films\cite{Ruzmetov2009,Jeong2013}, the sign of the Hall coefficient indicates the dominance of $n$-type carriers.  Interpreting the Hall data in terms of a single carrier type leads to the conclusion that there are several mobile carriers per vanadium ion.  This has resulted in the suggestion that both $n$- and $p$-type carriers are present in the system, leading to a comparatively small Hall number\cite{Rosevear1973,Ruzmetov2009}.  In both our as-grown and hydrogenated films, we observe a very small Hall signal at room temperature (Fig. 1e, hydrogenated film data) with a negative slope, consistent with a dominant contribution of $n$-type carriers.  For the hydrogenated films, the Hall voltage becomes larger as the temperature decreases, which conventionally implies a decrease in carrier density. Linear fi
 ts are performed to obtain the slope of the Hall data.

Although the very low Hall resistance implies that the single-carrier picture likely does not fully describe the transport in metallic VO$_2$ (either the high-temperature state of the pristine material, or the hydrogen-doped conducting state), there is not sufficient information available in the Hall and longitudinal resistivity data to extract  separate carrier density and mobility information for two carrier types. We thus follow the historical precedent of analysis under the (majority) $n$-type single carrier picture. We are able to deduce the Hall carrier density $n$ and mobility $\mu$ from the Hall measurements and the resistivity, according to the equations: $n=IB/(V_{\mathrm{H}} ed)$, $\mu =1/en\rho$ , where $I$ is the current, $B$ is the perpendicular magnetic field, $V_{\mathrm{H}}$ is the Hall voltage, $d$ is the thickness of the sample, and $\rho$ is the resistivity.  The temperature dependence of the deduced $n$ and $\mu$ is presented in Fig. 1f.  The extremely sm
 all Hall voltage leads to an extraordinarily high calculated carrier density, further resulting in an extremely small calculated value of mobility.  At room temperature, the deduced $n$ for the 10~nm film sample HP1704 is about $45 \times 10^{28}$~m$^{-3}$, much larger than the density of vanadium atoms ($\sim 3 \times 10^{28}$~m$^{-3}$).  Even at 2~K, where $n$ reaches the minimum in the limit of our measurement setup, the inferred value is still as large as $1.9 \times 10^{28}$~m$^{-3}$.

   All these results further support the conjecture originating in the pristine VO$_{2}$ literature that the assumption of a single carrier type is incorrect; thus both $n$- and $p$-type carriers are likely present in this system, with comparable contributions to the conductance.  While a two-carrier interpretation has been long been suggested to explain the small Hall number (and therefore large inferred carrier density) in metallic VO$_2$,\cite{Berglund1969} there remains no direct evidence for this, or an explicit electronic structure treatment that clearly identifies a possible source of hole carriers.  This aspect of the band structure ($e.g.$, the source of the small Hall number; any hole-like source of carriers that could contribute) of doped (or even pristine metallic) VO$_{2}$ remains unresolved.  

\begin{figure*}[!htb]
\includegraphics[width=\textwidth]{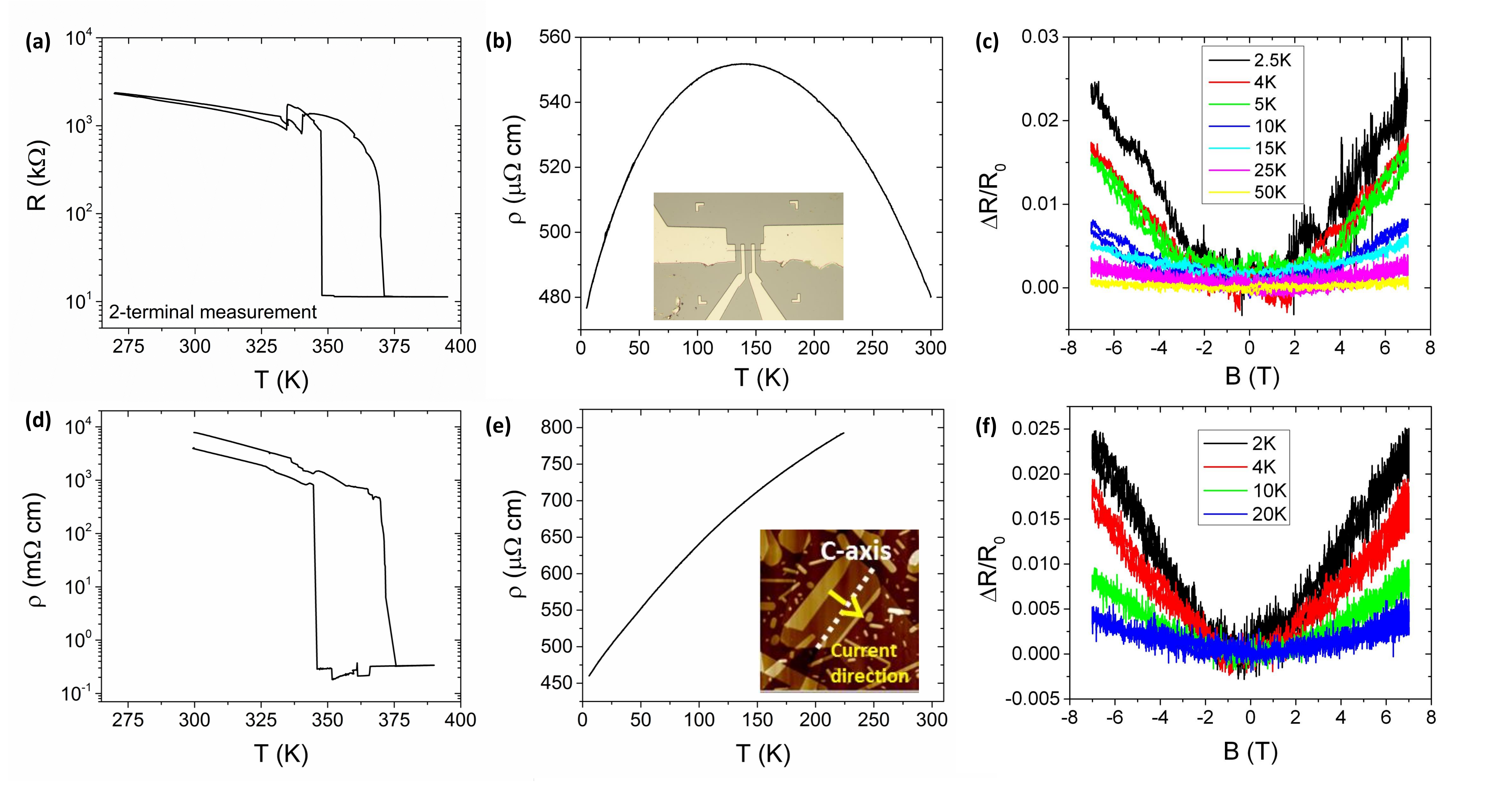}
\caption{ Top row: Temperature dependence of the resistance of a VO$_{2}$ nanobeam sample (a) before and (b) temperature dependence of resistivity after hydrogenation. The inset of (b) is a photo of the measured nanobeam test structure, taken under an optical microscope. (c)  Positive MR response of hydrogenated VO$_{2}$ nanobeam sample at various temperatures with the field perpendicular to the substrate plane. Bottom row: Temperature dependence of resistivity of a VO$_{2}$ micron-size flake like sample (d) before and (e) after hydrogenation, with the inset of (e) showing the atomic force microscopy (AFM) image of the flake sample measured. Panel (f) shows its MR responses at various temperatures with the field perpendicular to the substrate plane.\label{fig:figure3}}
\end{figure*}

\subsection{VO$_{2}$ nanobeams}

As a comparison, the temperature dependence of resistance and the magnetoresistance of hydrogenated VO$_{2}$ nanobeams were also measured under comparable temperature and field conditions.  The resistance of a representative sample shows a non-monotonic temperature dependence (Fig. 3(b)). The resistance gradually increases as the temperature decreases from 300~K to 140~K, at which point it reaches a maximum, though the total change is less than 20\% and is thus quite different from the temperature dependence of resistance of a semiconductor.  Below 140~K, the resistance decreases with cooling, similar to the behavior of a metal.  Based on the temperature dependence of resistance measurement of the pristine sample, this transferred nanobeam should be rutile under $c$-axis tensile strain, which is due to direct contact with the substrate. When grown at high temperatures, the undoped nanobeams and flakes are initially in the rutile state, and are strongly elastically coupled to 
 the underlying SiO$_2$ layer.  After cooling to room temperature, the nanobeams transition into the monoclinic state (with its differing lattice parameters), yet remain firmly mechanically clamped by the underlying SiO$_2$, and thus become compressively strained along the rutile $c$-direction due to differences in thermal expansion.  Nanobeams transferred onto a non-growth substrate are then in a comparatively unstrained state when they are in the monoclinic insulating phase, but they tend to be adhered to the substrate by van der Waals forces; when they transition to the metallic phase (either by heating or hydrogen doping), they will be under tensile strain. This is in contrast to the case of a VO$_{2}$ film grown on a TiO$_{2}$ substrate, which has a much stronger rutile $c$-axis compressive epitaxial strain due to the lattice mismatch\cite{Muraoka2002}. 

The magnetoresistance of the VO$_{2}$ nanobeam shown in Fig. 3(c) has a positive high-field MR response below $\sim25$~K, similar to the films, but without evidence of a low-field peak or dip feature at the lowest temperatures.  Since the width and thickness of the nanobeam are both more than 50~nm, and the MR measurement with the magnetic field parallel to the current indeed shows the same rough magnitude of change as the perpendicular case (see Fig. S3), it seems reasonable to conclude that this system is also effectively 3d from the perspective of localization corrections.  The appearance of positive MR at low temperature here and in the films suggests the possibility of weak anti-localization, and implies that vanadium may possess strong enough spin-orbit scattering to observe WAL.  This is not entirely surprising, considering that some positive weak anti-localization MR is observed in copper films\cite{Bergmann1984} whose atomic number ($Z=29$) is only slightly larger th
 an that of vanadium ($Z=23$).  As in the films, the MIRL resistivity here precludes a self-consistent quantitative analysis of the MR in terms of WL/WAL.

Even though the positive MR of both VO$_{2}$ films and nanobeams may be interpreted as a consequence of weak anti-localization, the reason for the differences in the low-field MR response of the two sample types remains unclear.  Beyond the question of the exact hydrogen concentration, there are two significant differences between these kinds of samples: strain direction and current direction with respect to the $c_{\mathrm{R}}$-axis.  The difference of strain directions has been mentioned above, and is clearly important based on the fact that the phase transition temperature in the undoped samples shifts to opposite directions in these two morphologies.  It is long established that strain can modify the spin-orbit coupling\cite{Liu1962}; in principle a strain-induced modification of the band structure and inversion symmetry could explain a sign flip in weak (anti)localization MR if the resulting SOC is sufficient.  An example of strain modifying SOC and in turn the MR is Hab
 ib et al.\cite{Habib2007}, where SOC is explicitly tuned in GaAs 2d holes as a function of strain. Without a detailed electronic structure calculation, however, that explains, $e.g.$, the small Hall number in undoped metallic VO$_{2}$ and the electronic structure of the hydrogen-stabilized phase, or further spin-sensitive experiments, it is difficult to draw firm conclusions about the interplay of strain and SOC in this system.

The current direction may also be important if the hydrogenated VO$_{2}$ has anisotropic electronic properties.  The physical structure (one-dimensional V chains in the undoped rutile structure) and highly anisotropic H diffusion properties clearly show that the $c_{\mathrm{R}}$-axis is a special direction in VO$_{2}$; it is the direction along which dimerization of the V occurs in the insulating state of VO$_{2}$.  For films, the current is in-plane, and the $c_{\mathrm{R}}$-axis is perpendicular to the plane, therefore, the current is perpendicular to the $c_{\mathrm{R}}$-axis.  In nanobeams, however, the current is by necessity parallel to the $c_{\mathrm{R}}$-axis.  To investigate the possible existence of anisotropy, we turn to measurements of VO$_{2}$ flake-like single-crystal samples.

\subsection{VO$_{2}$ microflakes}

The $c_{\mathrm{R}}$-axis of the VO$_{2}$ micro flakes can be determined by optical microscopy, because the domains formed during the phase transition in the undoped material are always in a stripe shape perpendicular to the $c_{\mathrm{R}}$-axis, and are visible through their optical contrast. Generally, the $c_{\mathrm{R}}$-axis is along the long side of rectangular VO$_{2}$ flakes.  Here, four V/Au contacts were deposited along the short side of the flake, so that the current would be perpendicular to the $c_{\mathrm{R}}$-axis (inset of Fig. 3(e)), as in the film measurements, and in contrast to the nanobeam measurements.  Although the VO$_{2}$ flakes remain on their growth substrate, and thus are expected to have larger strain than transferred VO$_{2}$ nanobeams, the hydrogenated flake sample shows a monotonic positive temperature-dependent resistivity (Fig. 3(e)), similar to a metal.

The MR measurements of the flake sample show a positive response that grows with decreasing temperature (Fig. 3 (f)), just as for the nanobeam case.  Since the current direction is now perpendicular to the $c_{\mathrm{R}}$-axis, this essentially rules out the possibility that the different MR response of the films and the single-crystal wires/flakes (the low-field peak or dip features seen in the films) is caused by intrinsic anisotropy.  

\begin{figure}[!htb]
\includegraphics[width=\textwidth]{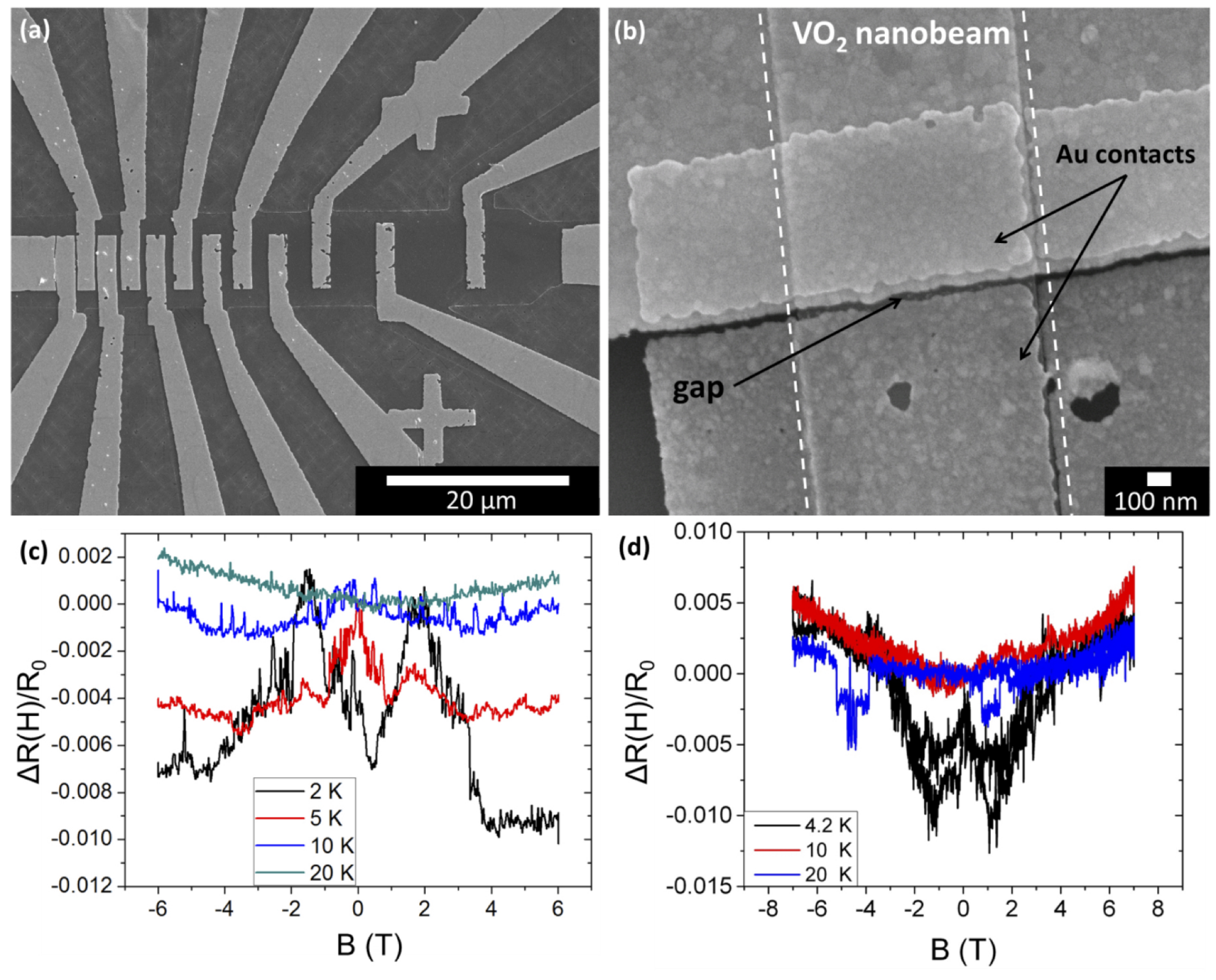}
\caption{(a) SEM image of a patterned and etched VO$_{2}$ film sample, with spacings between the gold contacts ranging from 300~nm to 8~$\mu$m. (b) SEM image of a VO$_{2}$ nanobeam sample patterned by a self-aligned technique, with a gap size of about 20~nm. (c) MR response of the hydrogenated VO$_{2}$ film with a contact spacing of about 400~nm. (d) MR response of the hydrogenated VO$_{2}$ nanobeam nano-gap sample. Both samples show apparently random, but retraceable MR fluctuations, consistent with UCF physics.\label{fig:figure4}}
\end{figure}

\subsection{Mesoscopic conductance fluctuations}

We also study submicron-size hydrogenated VO$_{2}$ test structures, consisting of both an epitaxial film sample and a nanobeam sample. On the film, we use e-beam lithography and e-beam evaporation to define and deposit multiple contacts with spacings between voltage probes varying from 500~nm to 8~$\mu$m, and then use e-beam lithography and reactive-ion etching (RIE) to pattern and etch away the unwanted VO$_{2}$ between the gold contacts except for a strip about 10~$\mu$m wide in the middle (Fig. 4(a)). The cracked 30~nm thin film is used for these measurements, though the length scales between the metal contacts are sufficiently small that the cracks are not expected to affect the transport. Qualitatively similar results are obtained on nanobeam samples as well.   For the nanobeam samples, a self-aligned technique is used to deposit gold contacts, forming a gap size on the order of 100~nm (Fig. 4(b)).  All these samples are hydrogenated using the same process described abov
 e, and transport measurements are done in the PPMS.  Although the magnitude of resistivity and its temperature dependence show behavior similar to that of the large-scale samples, the MR is quite different. In both film and wire samples with submicron voltage probe separations, the MR varies nonmonotonically in an apparently random manner (Figs. 4(c) and 4(d)), with some signs of telegraph noise.  The randomness, however, still follows certain rules: the curves retrace themselves with increasing or decreasing magnetic field; are approximately symmetric (to within the effects of telegraph switching) about  zero field; and the lower the temperature, the larger the fluctuations.  For a given sample, after warming up to room temperature and cooling down again, the pattern of MR changes.  All these features suggest that the fluctuations are likely a form of universal conductance fluctuation (UCF)\cite{Lee1985}.  The large field scale associated with the observed fluctuations is c
 onsistent with the short coherence length on the order of $\sim$10~nm implied by 3d WLMR.  If the MR fluctuations are interpreted in terms of a fluctuating parallel conductance, the change of conductance is less than 1~$G_{0}$ ($2e^{2}/h \sim 7.748 \times 10^{-5}$~S), consistent with expectations of UCF, which is suppressed with increasing sample size through ensemble averaging.  The large field scale precluded detailed studies of the conductance fluctuations ($e.g.$, autocorrelation of the conductance, $G$, as a function of magnetic field).

\section{Conclusions}

In this work, we studied the low temperature transport properties of hydrogenated VO$_{2}$ of different morphologies that were grown by distinct methods.  The large resistivity of these samples apparently violates the Mott-Ioffe-Regel limit over the full temperature range examined, implying the failure of simple semiclassical transport for this doping-stabilized metallic state of H$_{x}$VO$_{2}$.  VO$_{2}$ films grown by MBE on TiO$_{2}$ (001) substrates show a positive high-field MR when $T < 150$~K independent of field direction.  The detailed low-field MR structure shows either a dip or peak at low temperatures, with the sign likely determined by a combination of strain effects and the hydrogen doping level and uniformity. Analogous MR measurements of VO$_{2}$ nanobeams and micron-scale flake-like crystals grown by PVD on Si/SiO$_{2}$ substrates show a high-field positive response similar to that of the films, but with a low-field response that is featureless (no sharp pea
 k or dip) to within the noise background. This positive overall MR resembles a quantum interference correction to conduction like WAL/WL.  Quantitative analysis of the MR, however, is complicated by poor understanding of the electronic structure of the bad metal state and a lack of a self-consistent theoretical framework for analyzing such quantum interference effects in MIRL-violating materials. Hall measurements on the film material confirm a very small Hall number, both in the pristine room-temperature metallic state and in the hydrogen-stabilized metallic state.  Finally, magnetoresistance measurements on sub-micron scales show evidence for mesoscopic conductance fluctuations with a field scale consistent with a short ($\sim 10$~nm) coherence length.

These observations demonstrate that mesoscopic corrections to electronic conduction can be present even in materials that apparently violate the Mott-Ioffe-Regel limit, and hence should conventionally be expected to be crossing into the strongly localized regime.  Our experiments highlight the need for a better understanding of the electronic properties of metallic VO$_{2}$ (undoped or stabilized by hydrogen doping), and the importance of developing treatments of mesoscopic effects in systems where the conventional, long-useful, semiclassical picture is not appropriate.

\section{Acknowledgements}
W.J.H. and H.J. contributed equally to this work. W.J.H., H.J, and D.N. gratefully acknowledge support from the US DOE Office of Science/Basic Energy Sciences award DE-FG02-06ER46337.  H.P. and D.G.S. thank the Center for Low Energy Systems Technology (LEAST), one of the six SRC STARnet Centers, sponsored by MARCO and DARPA for their financial support.  This work was performed in part at the Cornell Nanoscale Facility, a member of the National Nanotechnology Infrastructure Network, which is supported by the National Science Foundation (Grant No. ECCS-0335765). We thank Nam Dong Kim (Rice University) for assistance with atomic hydrogenation.

 ffset by (c) -0.002 or (d) -0.01  for clarity. An overall high-field  positive MR develops as the temperature is reduced below $\sim$ 150~K, growing as the temperature decreases. A small dip (perpendicular case) or peak (parallel case) appears at 4~K and below, likely due to localization effects. (e) Hall resistance of the hydrogenated film measured at various temperatures. The data have been antisymmetrized to remove a longitudinal component artifact. (f) Calculated mobility (left axis) and carrier density (right axis) inferred from resisitvity and Hall data assuming a single carrier type ($n$-type). The unphysically large inferred carrier density implies that more than one carrier type likely contributes to transport. 

\newcommand{\beginsupplement}{%
        \setcounter{table}{0}
        \renewcommand{\thetable}{S\arabic{table}}%
        \setcounter{figure}{0}
        \renewcommand{\thefigure}{S\arabic{figure}}%
     }
\beginsupplement

\begin{figure*}
\includegraphics[width=\textwidth]{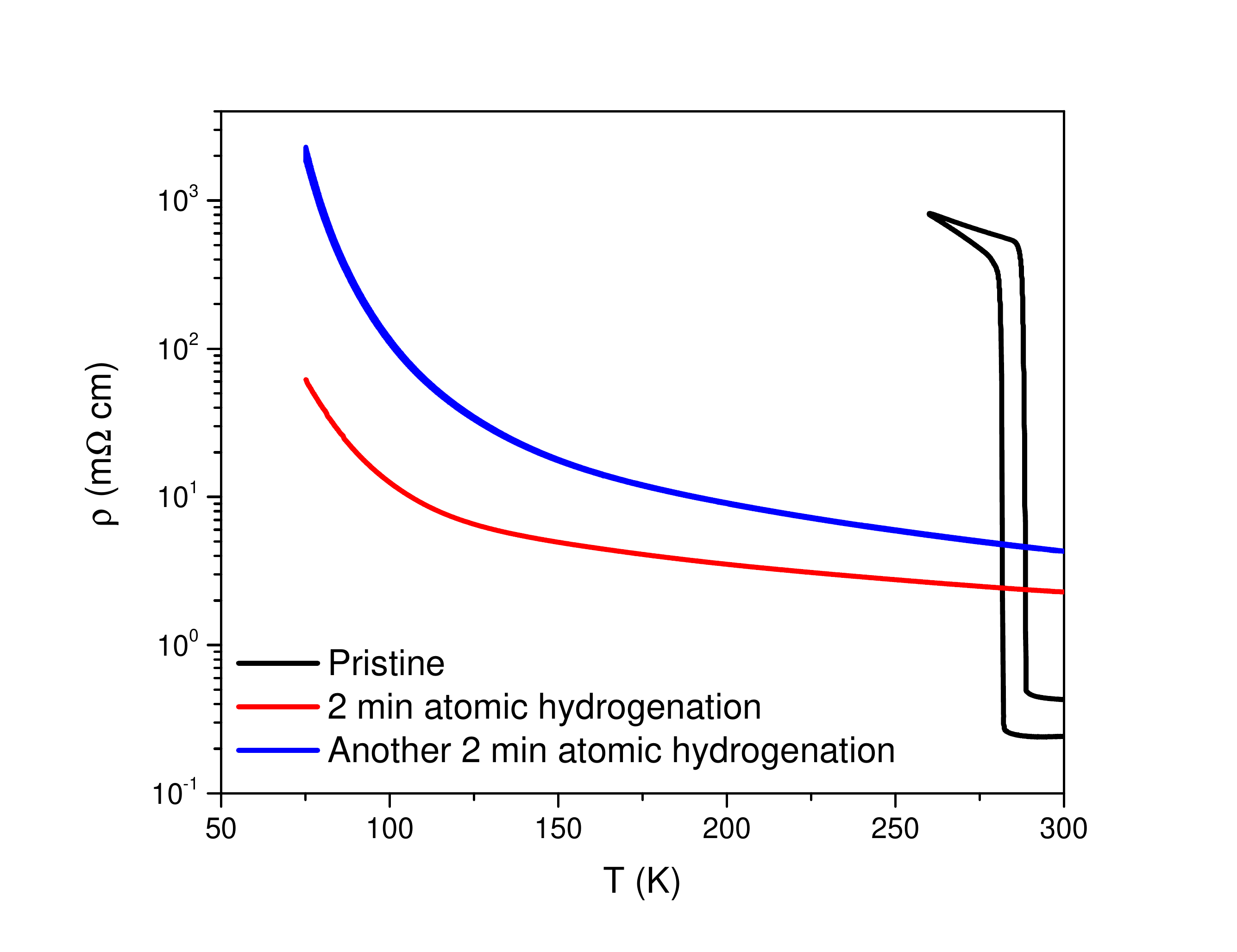}
\caption{Temperature dependent resistivity of film sample HP1703 (10~nm thickness), showing the effect of increasing atomic hydrogenation time. A two-minute hydrogen exposure results in an enhanced room-temperature resistivity (red curve) compared to the pristine state (black curve), and a second exposure of two minutes further increases according to this trend (blue curve). The insulating-like curve slope with decreasing temperature is much steeper and larger in magnitude than for film samples HP1704 and HP1706, which were doped for just 30s each. Such a trend toward insulating behavior with longer hydrogen exposure time was not observed in the PVD-grown nanobeams or flakes, likely due to the significantly slower diffusion time and lower hydrogen doping level for those samples compared to films (which is, in turn, a consequence of the different sample geometries and their crystal orientations and strain states).\label{fig:figureS1}}
\end{figure*}

\begin{figure*}
\includegraphics[width=\textwidth]{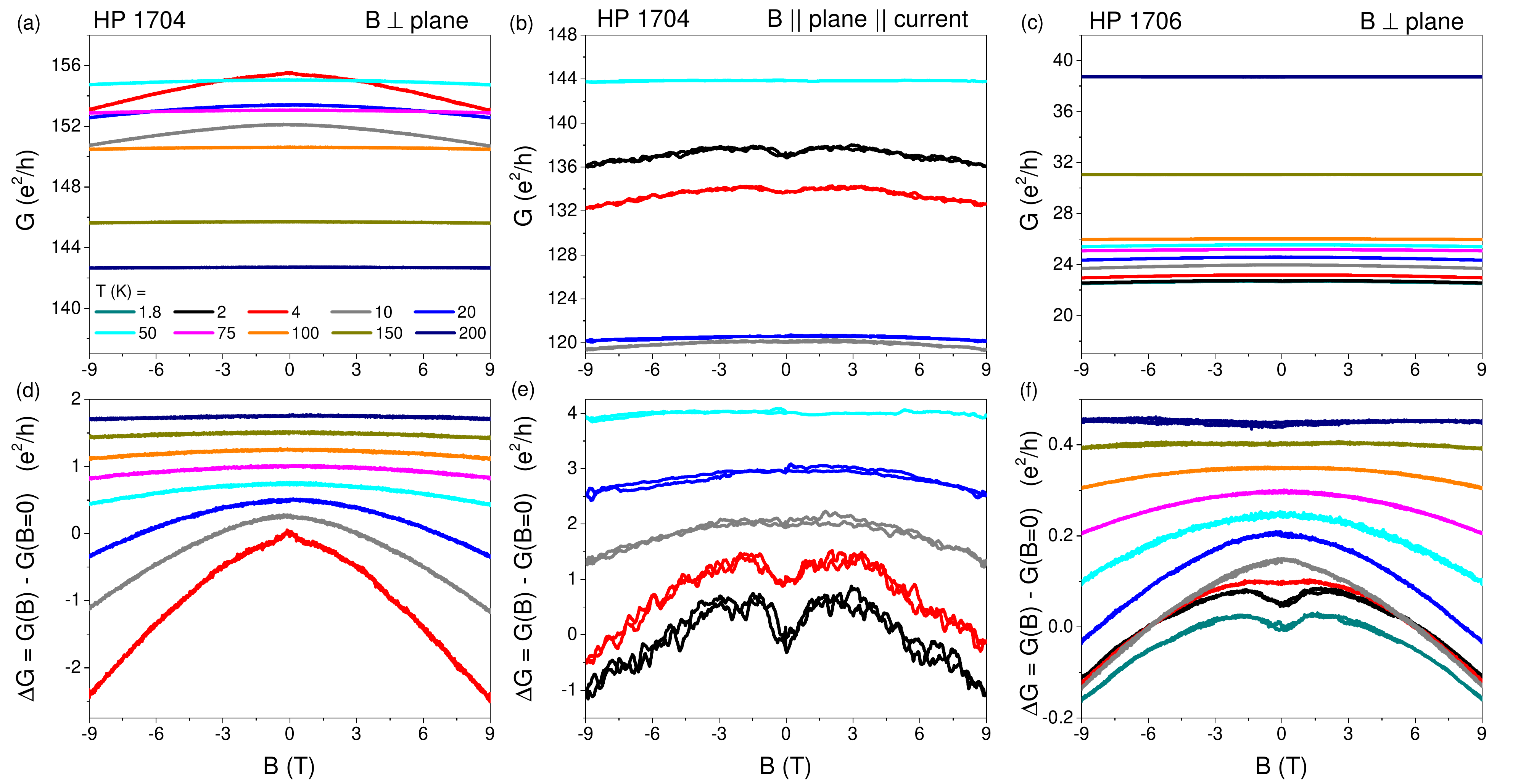}
\caption{Magnetoconductance for film sample HP1704 with the field (a,d) perpendicular or (b,e) parallel to the plane of the substrate, and (c,f) film sample HP1706 with the field perpendicular to the plane of the substrate. The top row shows plots in units of raw conductance, and the bottom row shows the difference $\Delta G = G(B)-G(B = 0)$. Curves in the bottom row have been offset for clarity by (d) 0.25, (e) 1, and (f) 0.05. \label{fig:figureS2}}
\end{figure*}

\begin{figure*}
\includegraphics[width=\textwidth]{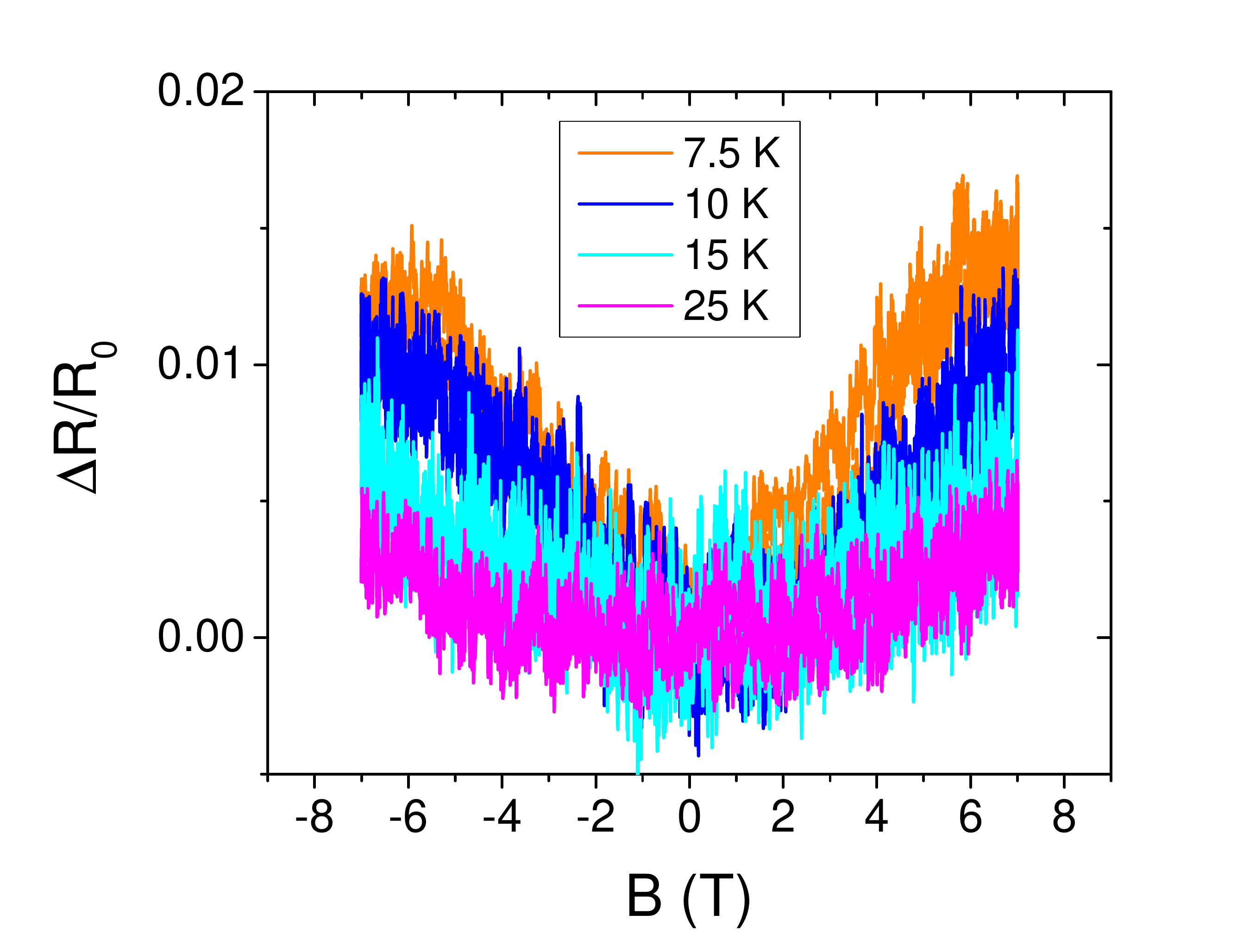}
\caption{Positive MR response at various temperatures of the same hydrogenated VO$_2$ nanobeam sample as in Fig. 3(a-c), with the magnetic field oriented in the plane of the substrate and parallel to the current. The MR shape and magnitude are similar to what was observed in the field-perpendicular-to-plane orientation at comparable temperatures. \label{fig:figureS3}}
\end{figure*}

\end{document}